\documentclass[prc,twocolumn,showpacs]{revtex4}
\usepackage{amsmath}
\usepackage{graphicx}
\usepackage{float}

\begin{document}

\title{Elliptic flow in high multiplicity proton-proton collisions at $\sqrt{
\text{s}} = $14~TeV as a signature of deconfinement and quantum energy
density fluctuations.}
\date{\today}
\author{G. Ortona}
\affiliation{Universit\`a di Torino \& Istituto Nazionale di Fisica Nucleare,
Torino, Italy}
\author{G.S. Denicol}
\affiliation{Institute f\"ur Theoretische Physik, 
Johann Wolfgang Goethe-Universit\"at, Max-von-Laue Str. 1,
60438, Frankfurt am Main, Germany}
\author{Ph. Mota}
\affiliation{Instituto de F\'isica, Universidade Federal do Rio de Janeiro, Rio de
Janeiro, Brazil}
\author{T. Kodama}
\affiliation{Instituto de F\'isica, Universidade Federal do Rio de Janeiro, Rio de
Janeiro, Brazil}

\begin{abstract}
At LHC extreme values of energy density will be reached even for
proton-proton collisions. Such values of energy density may be large enough
to generate a collective motion in the products of the collision, therefore
generating effects such as elliptic flow. Using ideal 3+1D hydrodynamical
simulations, we show that elliptic flow can occur at least for top
multiplicities p-p events at LHC and that the intensity of such effect is
strongly related to quantum fluctuations in the initial proton energy
distribution.
\end{abstract}

\maketitle

\section*{Introduction}

The application of hydrodynamical models to describe hadronic collisions
have a long history, dating almost from the beginning of the studies of
hadronic interactions\cite{Fermi,Landau,Carruthers}. Since late '80s these
approaches have been revived to describe the collective features of the
dynamics in heavy ion collisions at the big accelerators\cite%
{Strottman,Csernai,Stocker} and investigate the properties of quark-gluon
plasma (QGP). The main idea behind such approach is that from the analysis
of collective flow in the dynamics, if any, we may infer the properties of
the allegedly deconfined partonic matter generated in heavy ion collisions.
For this, the thermalization of the partonic matter should occur in a
relatively short time and space scales. Until today this approach was
limited to heavy ions collisions because only in these collisions a large
space-time domain, in which the energy density is sufficiently high to lead
to the deconfinement and thermalization, is expected to emerge.
Proton-proton (p-p) data were used as reference to analyze these heavy ion
data. 

In p-p processes the system is considered too small and, so far, it was not
expected to get any deconfinement nor thermalization domains. Thus, any
differences in behavior between the p-p and heavy ions case may (should) be
attributed to the dense medium effects.

As mentioned above, one of the most important signatures of the creation of
such a medium is the manifestation of collective motions (\emph{flows}) in
the final state particles. The presence of any collective flow is a typical
behaviour of a fluid matter. There are many ways to identify collective
flows. They appear as an anisotropy in angular distribution of particles and
are referred to as anisotropic flow. In experimental studies, anisotropic
flows are usually analyzed in terms of the Fourier expansion of the
azimuthal particle distribution 
\begin{equation}
\frac{1}{2\pi N}\frac{dN}{d\phi}=1+2v_{1}\cos(\phi)+2v_{2}\cos\left( {2\phi }%
\right) +...
\end{equation}
where usually $v_{1}$ is referred as \emph{direct flow} and $v_{2}$ as \emph{%
elliptic flow.}\newline
It should be noted that these anisotropic parameters are not necessarily
null even for free-streaming particles, but their dependence on quantities
such as, for example, impact parameter and kinematic variables will reflect
the collective features of the dynamics.

LHC accelerator, that is starting in Geneva, will reach extreme values of
energy both for heavy ions and protons, the latter being accelerated up to $%
\sqrt{s}=14$~TeV. It is possible that under these extreme conditions, also
in proton-proton collisions the energy density will reach values high enough
to allow partonic matter to survive in a deconfined state for a time long
enough to thermalize and show some collective behaviour. In particular, when
we consider quantum fluctuations in the incident states, such a situation
may well occur and thus manifest in large multiplicity events. In this
sense, collective features, if any, of the high multiplicity events may
reveal interesting informations on the initial condition for high energy p-p
collisions.

The goal of this work is to estimate the elliptic flow for top multiplicity
proton-proton events in the framework of ideal hydrodynamics and investigate
if it will be possible to perform such analysis at the LHC. If viscosity is
present, it has effects on collective flow and to determine the viscosity of
the QGP fluid is presently one of the important problems. However, these
questions are still under investigation, and in the present work, we only
consider the case of an ideal fluid to avoid additional ambiguities
associated to viscous theories\cite{bulkviscosity}.

\section{EoS and initial conditions}

For simplicity, we use a factorized initial energy density profile into its
longitudinal and transverse parts, as proposed by \cite{HiranoIC}. For the
transverse part, we use an energy density profile as a simple superposition
of those of colliding particles as is usually done for heavy ions collisions 
\cite{e0trans}. Thus we use the formula 
\begin{equation}
\begin{split}
\epsilon _{0}& (x,y,b)= \\
& k(\sqrt{s})T_{A}(x+\frac{b}{2},y)\left[ 1-(1-\frac{\sigma T_{B}(x-\frac{b}{%
2},y)}{B})^{B}\right] + \\
& k(\sqrt{s})T_{B}(x-\frac{b}{2},y)\left[ 1-(1-\frac{\sigma T_{A}(x+\frac{b}{%
2},y)}{A})^{A}\right] ,
\end{split}
\label{eq:etransAA}
\end{equation}%
where $A$ and $B$ are the number of nucleons, $\sigma $ is the
nucleon-nucleon cross section and $k$ a collision energy dependent constant.
In the case of proton-proton collisions all this is reduced to the simple
expression 
\begin{equation}
\epsilon _{t=0}(x,y,b)=2k\sigma T_{A}(x+\frac{b}{2},y)T_{B}(x-\frac{b}{2},y).
\label{eq:etranspp}
\end{equation}%
For the proton energy density $T_{A(B)}$, we use a Gaussian distribution
with a top value $\epsilon _{0}$ and width $\sigma _{p}~=~0.875$~fm. 
\begin{equation}
T_{A}(x,y)=\epsilon _{0}e^{-\frac{x^{2}+y^{2}}{2\sigma _{p}}}.
\label{eq:pdens}
\end{equation}%
As the convolution of two Gaussian is still a Gaussian, our initial
transverse energy density is 
\begin{equation}
\epsilon _{t=0}(x,y,b)=Ke^{-\frac{r^{2}}{\sigma _{p}}},  \label{eq:etrans}
\end{equation}%
where all the constants ($k,\sigma $...) are compacted together in the new
constant $K$. It must be noticed that the result of eq. (\ref{eq:etrans}) is
completely symmetric in the transverse plane for any value of $b$. We will
discuss this point in more detail later in section \ref{sec:conc}.

\begin{figure}[h]
\begin{center}
\includegraphics[height=5cm,width=\linewidth]{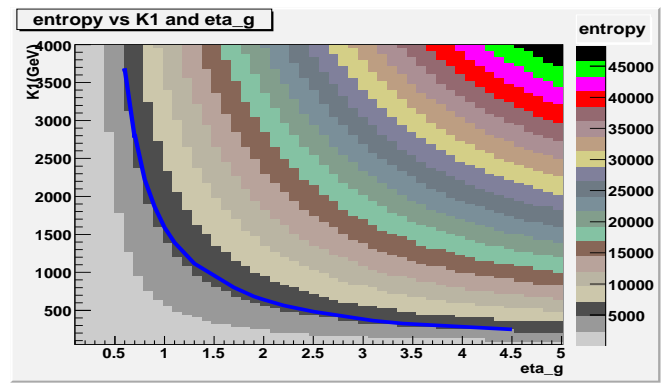}
\end{center}
\caption{(Color online) entropy dependence on the parameters $\protect\eta %
_{g}$ and $K$. The solid line shows the entropy region reproducing the top
LHC multiplicities. }
\label{fig:entropia}
\end{figure}

For the longitudinal dimension, we proceed as in \cite{HiranoIC} using the
equation: 
\begin{equation}
\epsilon _{long}(\eta _{s})=e^{-\frac{\eta _{s}^{2}}{2\eta _{g}^{2}}}\theta
(y_{beam}-\eta _{s}).  \label{eq:elong}
\end{equation}%
We are left with two free parameters, the constant $K$ and the longitudinal
width $\eta _{g}$. We can express the total entropy as a function of these
parameters, and so relate them to the particle multiplicity. as shown in
Fig. (\ref{fig:entropia}). Remarkably, the results on the integrated value
of $v_{2}$ are almost not affected by the choice of these parameters,
provided that they keep the total entropy constant. Furthermore, it depends
explicitly on the shape of the impact region, as is discussed later.
Nonetheless, the momentum dependence of $v_{2}(p_{t})$ is strongly affected
by those parameters (cf. sec. \ref{sec:results}).

We have chosen entropy values between 5400--5500~fm$^{-3}$ in order to
reproduce the top multiplicities expected at LHC (HIJING simulations) for
such collisions. It is very important to focus on very high multiplicity
events for two reasons. The first is that proton-proton collisions on
average have very low multiplicity. At LHC energies for proton-proton
collision we expect $dN/dy\approx 6$ while the top multiplicity can be
larger then one thousand particles. To extract any information on
\textquotedblleft collectivity\textquotedblright , the events producing a
large enough amount of particles are interesting. The second reasons is that
in low multiplicity events also non-flow correlations may become strong to
override the elliptic flow, even if it exists. 
Fixing the total multiplicity, we further determine the parameter $\eta _{g}$
by the expected mid-rapidity multiplicity of particles. This choice leads to
the values $\eta _{g}$~=~1.3 and $K$~=~1150~GeV. In order to test the
momentum dependence of $v_{2}$ on the parameter choice, other two sets of
parameters are used in the paper: the first one ($K=370$~GeV and $\eta
_{g}=3.1 $) with very low top energy density but extremely broad
distribution in the $\eta \tau $ dimension, and the second one ($K=3680$~GeV
and $\eta _{g}=0.6$) with a very high initial energy density but almost
confined in the transverse plane.

In ideal hydrodynamics the elliptic flow is proportional to the initial
eccentricity of the impact region \cite{Ollitrault}.\newline
In our case, there is a big difference from heavy ion collisions. Due to the
Gaussian shape of the proton energy densities, the total energy density
profile eq.(\ref{eq:etranspp}) is symmetric, so that at the classical level
we would not expect elliptic flow at all. It has already been shown that
quantum effects on energy distribution are not much effective on large
systems, such as Au-Au collisions \cite{b0AuAu} but their importance grows
as the size of the system is reduced, like for Cu-Cu collisions \cite{b0CuCu}%
. So for the very small systems expected in p-p collisions, quantum
fluctuations should have a crucial effect on the event-by-event energy
density distribution of the created matter. We try to mimic such quantum
fluctuations by simply cutting the shape of the interaction region with an
elliptic shape parameterized by eccentricity $e$. We thus suppose that the
initial energy density for the hydrodynamical evolution as 
\begin{equation*}
\varepsilon (\vec{r},\eta ,e)=Ke^{-\frac{r_{T}^{2}}{\sigma _{P}^{2}}}e^{-%
\frac{\eta _{s}^{2}}{2\eta _{g}^{2}}}\theta (y_{beam}-\eta _{s})\theta
\left( 1-\frac{x^{2}}{a^{2}}-\frac{y^{2}}{b^{2}}\right) ,
\end{equation*}%
with%
\begin{eqnarray*}
a &=&\sqrt{\frac{R}{2}(1-e)}, \\
b &=&\sqrt{\frac{R}{2}(1+e)}.
\end{eqnarray*}%
According to our hypothesis, the parameter $e$ measures the effect of
quantum fluctuation in the formation of the initial condition for ultra-high
energy p-p collisions.

As for the equation of state (EoS), we use the one shown in Fig.(\ref%
{fig:eos}) (solid line). This EoS is obtained by smoothly connecting the
equations of state of the lattice QCD result in \cite{aokiEOS} and an hadron
resonance gas near $T_{c}~=~200$~MeV. The phase transition is a cross-over.
To see the effect of the EoS on our results, we also test the EoS with a
steeper connection between these two phases (dashed). 
\begin{figure}[tbh]
\begin{center}
\includegraphics[height=4cm,width=\linewidth]{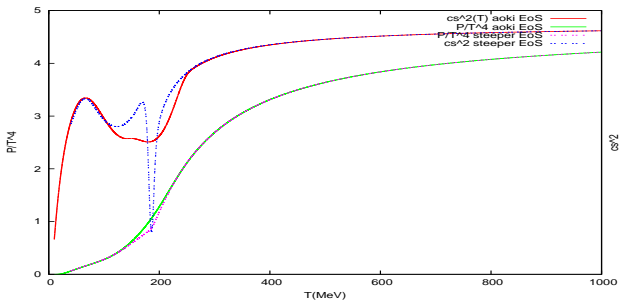}
\end{center}
\caption{(Color online) Behaviour of squared speed of sound and pressure
over $T^{4}$ as functions of temperature for the two different equations of
state.}
\label{fig:eos}
\end{figure}
After the initial entropy distribution is generated, the time evolution of
the system is followed using the Smoothed Particle Hydrodynamic (SPH)
formalism in hyperbolic coordinate system \cite{SPH}. To calculate the
elliptic flow, we perform the commonly used sudden freeze-out through the
Cooper-Frye procedure: when a fluid element crosses the surface defined by
the temperature $T=T_{f}$ all interactions are assumed small enough so that
all particles can be consider as free. The freeze-out temperature is settled
at the value $T_{f}=130$~MeV. Tests with $T_{f}$ = 120~MeV shows that our
results are almost insensitive to the changes in the value of $T_{f}$. 

A full 3+1D ideal fluid dynamics using our SPH code for p-p requires at
least 30~000 total number of SPH particles with the kernel width $h=0.3$ fm 
\cite{SPH}. With this condition, one collision event consumes \symbol{126}8
hours of calculation with a desktop PC. The overall precision of calculation
is monitored by the conservation of the total energy and momentum which are
kept within the relative error of $10^{-2}$ during the whole time evolution.

\section{Results and Conclusions}

\label{sec:results}\label{sec:conc}

\begin{figure}[tbh]
\begin{center}
\includegraphics[height=4cm,width=\linewidth]{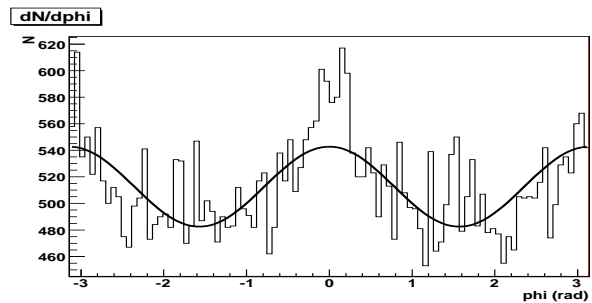}
\end{center}
\caption{Azimuthal thermal pions distribution for 100 high multiplicity
events. Fit leads to $v_{2}\approx 3\%$.}
\label{fig:v2test3100}
\end{figure}
\begin{figure}[htb]
\begin{center}
\includegraphics[height=4cm,width=\linewidth]{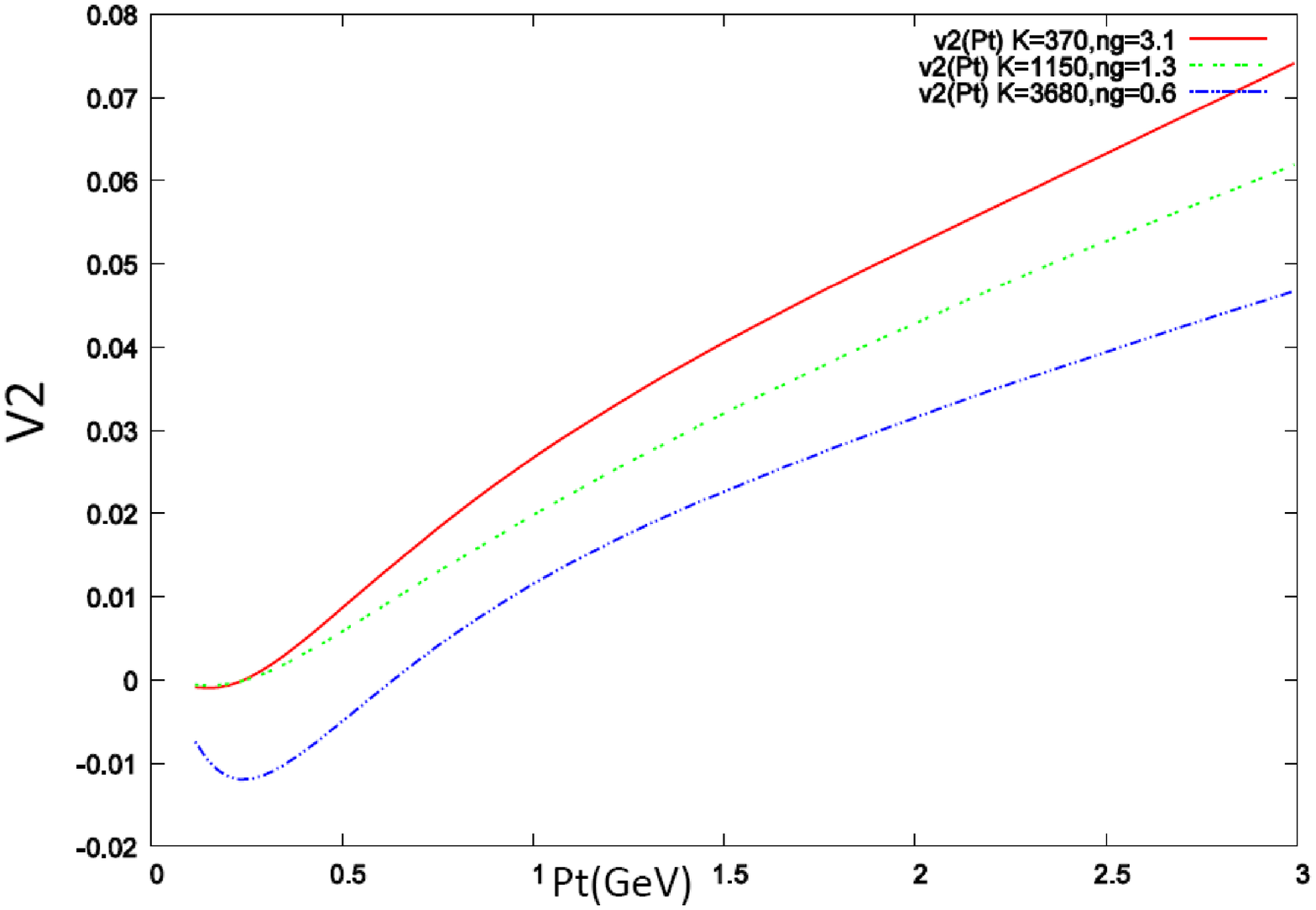} %
\includegraphics[height=4cm,width=\linewidth]{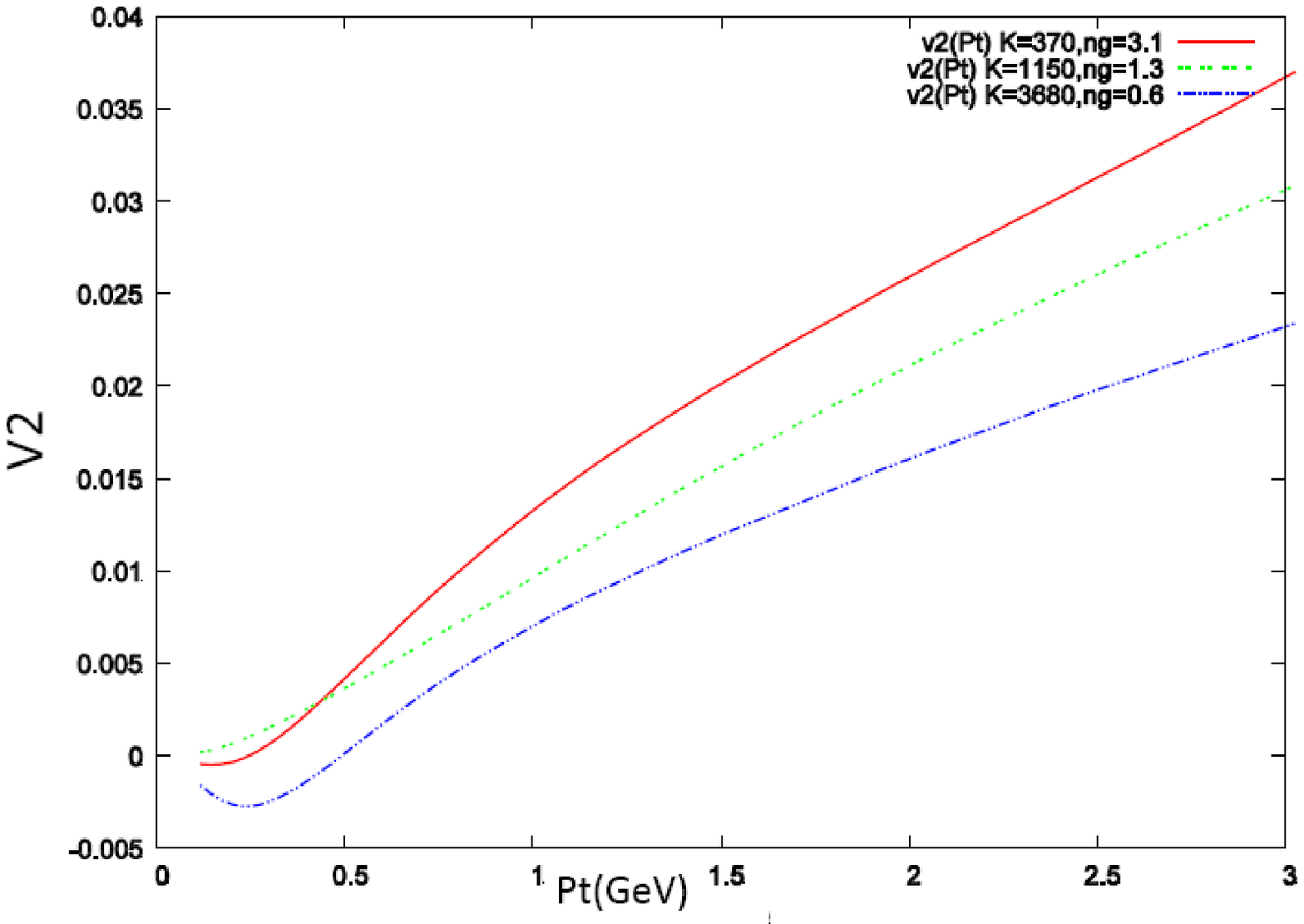} %
\includegraphics[height=4cm,width=\linewidth]{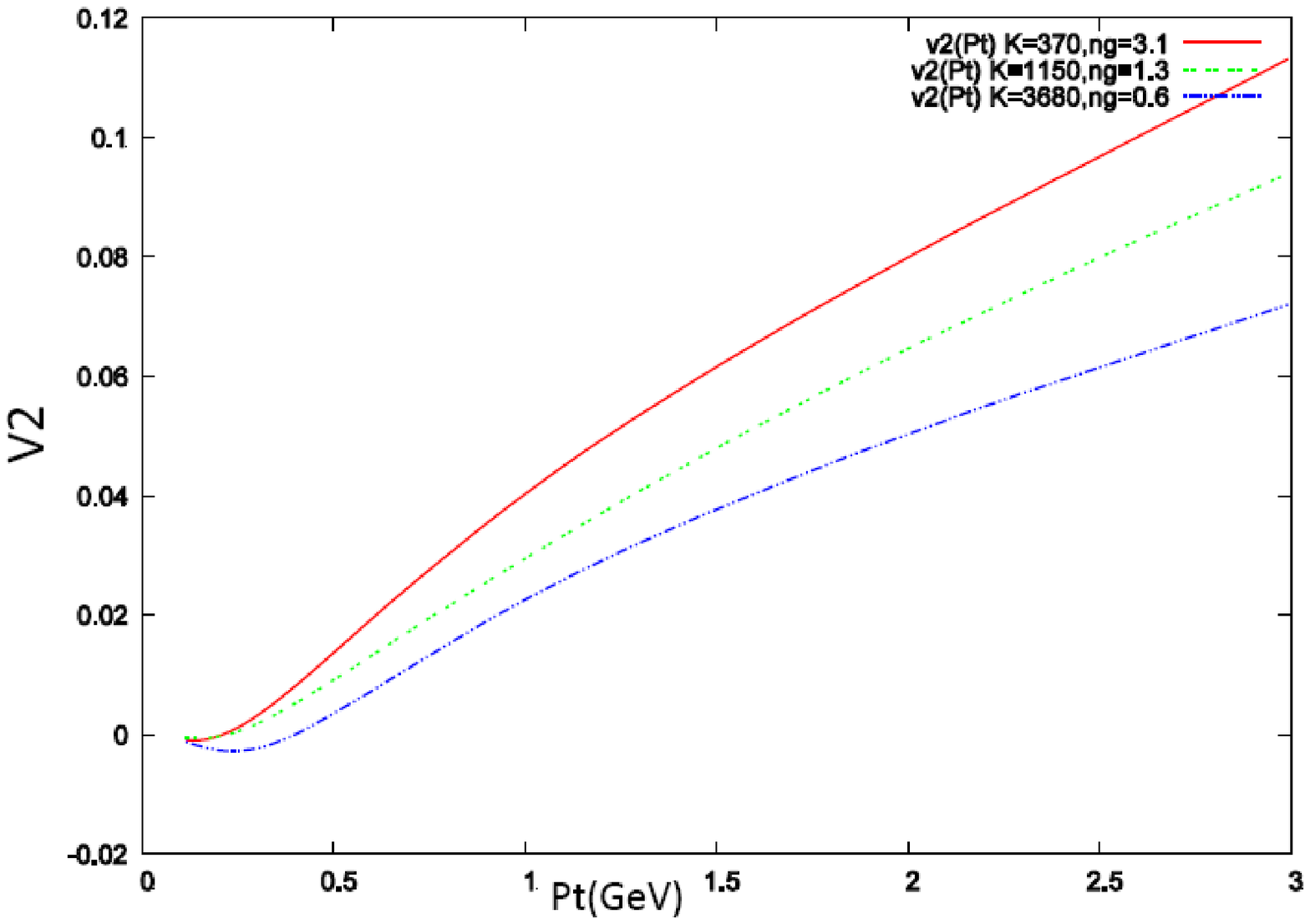}
\end{center}
\caption{(Color online) $v_{2}(p_{t})$ for an initial eccentricity of 17\%
(up), 8.6\% (center) and 26\% (bottom) for the three different set of
parameters}
\label{fig:v2b03}
\end{figure}

In Fig. (\ref{fig:v2test3100}) we present the azimuthal distribution of
thermal pions in the mid rapidity region $y\in \lbrack -1,1]$ for an initial
eccentricity of 17\%. The results are consistent with an elliptic flow of
2.9\%, so this indicate that an elliptic flow effect could be strong enough
to be experimentally measurable at LHC. In Fig. (\ref{fig:v2b03}) we show
the effect of the choice of the parameters on $v_{2}(p_{t})$ for different
values of initial eccentricity. As expected, the elliptic flow effect gets
stronger when most of the energy density is available in the transverse
plane and the integrated value of $v_{2}$ is proportional to the initial
eccentricity \cite{Ollitrault}. We also verify that the behavior of $v_{2}$
integrated over the whole rapidity domain is almost the same as that of the
mid-rapidity region. 
If the eccentricity $e$ has a quantum origin in p-p collisions, we expect
big fluctuations for the average value of $v_{2}$ for different multiplicity
bins in an experiment. Also, we should expect big fluctuations even in the
same bin for different events, as the origin of the eccentricity is not
geometrical but stochastic. We also expect the elliptic flow value to be
always greater than $0$, as any fluctuation in any collision contributes
with a positive value of $v_{2}$. However, the collective flow can only
become important for large multiplicity events. From our study, we get the
elliptic flow of $v_{2}\approx 3\%$ if the induced eccentricity has average
value of about 17\%. We expect that, measurements of elliptic flow in top
multiplicity events in p-p may give an important clue for the quantum
fluctuation in energy density distribution at each event.\newline
The ALICE collaboration is developing several methods to estimate non-flow
contribution to $v_2$ in p-p collisions\cite{prasad}. The most promising one
seems to be the $\eta$-gap method, that consists in using the particles with
pseudorapidity $\eta>3$ to evaluate the reaction plane and those at $|\eta|<1
$ to evaluate $v_{2}$. Other methods are, for example, the standard
event-plane method\cite{poskvol} or defining the reaction plane in the
direction of the leading particle at high pseudorapidity. Once non-flow
correlations are removed, $v_{2}$ can be measured. Unfortunately, there
aren't yet any available estimations of the precision such methods can reach
in removing non-flow correlations, but it is plausible that an effect of a
few percent would be visible in ALICE.\newline
As mentioned in the introduction, one thing that can affect the $v_{2}$
value is the presence of viscosity in the collective motion, that has been
neglected in our calculations. High value of viscosity can strongly affect
the value of the elliptic flow produced in proton-proton collisions, as
suggested in \cite{romats}. \newline
\begin{figure}[hbt]
\begin{center}
\includegraphics[height=4cm,width=\linewidth]{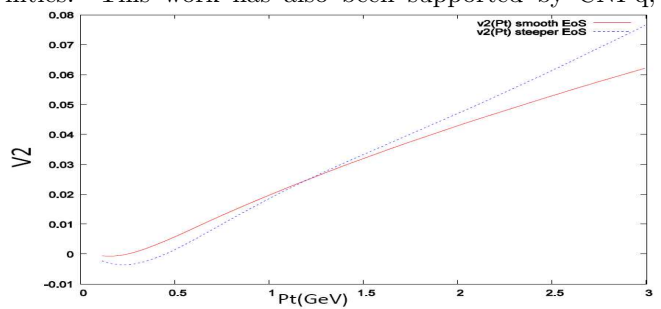}
\end{center}
\caption{(Color online) $v_{2}(p_{t})$ for the two different EoS presented
in fig. ( \protect\ref{fig:eos}) for eccentricity 17\%, K=1150~MeV and $%
\protect\eta _{g}=1.3$. }
\label{fig:eosb03}
\end{figure}

\section*{Acknowledgments}

We thank valuable discussions with Dr.T. Koide. We're also very thankful to
Mr S. K. Prasad, who shared with us his knowledge about non-flow effect
removal in p-p collisions.\newline
This work has been made possible thanks to the HELEN-ALPHA project, an
exchange program between European and Latin-American high energy physics
communities. This work has also been supported by CNPq, FAPERJ, CAPES,
PRONEX and the Helmholtz International Center for FAIR within the framework
of the LOEWE program (Landesoffensive zur Entwicklung Wissenschaftlich-
Okonomischer Exzellenz) launched by the State of Hesse.

\end{document}